\documentclass[12pt]{article}

\usepackage{amsmath}
\usepackage{graphicx}
\usepackage{cases}

\usepackage{natbib}
\usepackage{psfrag}
\usepackage{amssymb}

\def\be{\begin{equation}}
\def\ee{\end{equation}}
\def\bea{\begin{eqnarray}}
\def\eea{\end{eqnarray}}
\def\bml{\begin{mathletters}}
\def\eml{\end{mathletters}}

\begin{document}

\title{Multiple adaptive substitutions during evolution in novel environments}

\author{Kavita Jain$^{\dagger},^{\S}$ and Sarada Seetharaman$^{\dagger}$\\\mbox{}\\$^{\dagger}$ Theoretical
  Sciences Unit and $^{\S}$Evolutionary and Organismal Biology
  Unit,\\Jawaharlal Nehru Centre for Advanced Scientific Research, \\ Jakkur P.O., Bangalore 560064, India}

\maketitle

\newpage

\noindent
Running head: Adaptive walk in novel environment
\bigskip


\bigskip

\noindent
Corresponding author: \\
Kavita Jain, \\
Theoretical Sciences Unit and Evolutionary and Organismal Biology
Unit,\\
Jawaharlal Nehru Centre for Advanced Scientific Research, \\ 
Jakkur P.O., Bangalore 560064, India. \\
\texttt{jain@jncasr.ac.in}

\bigskip
   
\noindent
\textbf{Abstract:} We consider an asexual population  
under strong selection-weak mutation conditions evolving on rugged 
fitness landscapes with many local fitness peaks. Unlike the previous
studies in which the initial fitness of the population is 
assumed to be high, here we start the adaptation process with a low 
fitness corresponding to a population in a stressful novel 
environment. For generic fitness distributions, using an
analytic argument we find that the  
average number of steps to a local optimum varies logarithmically with
the genotype sequence length and increases as the correlations among 
genotypic fitnesses increase. When the fitnesses are exponentially 
or uniformly distributed, using an evolution equation for the 
distribution of population fitness, we analytically calculate the
fitness distribution of fixed beneficial mutations and the walk length
distribution.  

\newpage

%============================================================================
%INTRODUCTION
%============================================================================
Adaptation is an evolutionary process during which a population 
improves its fitness by accumulating beneficial mutations. A 
population of genotypic sequences produces a 
suite of mutants and if better mutants become available, a maladapted 
population may acquire one of the beneficial mutations provided it
does not get lost due to genetic drift. The fitter population in turn
may acquire another advantageous mutation and the process goes on
until the supply 
of beneficial mutations gets exhausted. A number of models with
variable degrees of biological consistency have been 
proposed and investigated to understand the process of adaptation 
\citep{Miller:2011}. One of the simplest mathematical models  was
introduced by Gillespie in which beneficial mutations arise
sequentially and fix rapidly \citep{Gillespie:1991}. If the
 mutation rate is small and the selection coefficient is large (compared
 to the inverse population size), it is a good approximation to assume
 that only the one-step mutants are accessible at any time 
and the population is localised at a single genotype. Such a 
 monomorphic population performs an adaptive walk by moving uphill on
 a fitness landscape until no more beneficial mutations can be found.

In the last few years, much of the work on Gillespie's model has focused 
on the first step in the adaptation process. If the fitness of the 
wild type and its one-mutant neighbors are rank ordered with the fittest
sequence at the top, 
the well established theory of extremes of independent random variables
\citep{David:2003} can be  
exploited  to obtain useful information provided the wild type has
a high fitness (rank). For a moderately high ranked initial fitness,  
Orr calculated the expected rank at the first step  assuming
exponential-like fitness distributions 
\citep{Orr:2002}. His prediction has been tested in an  experiment using 
single-stranded DNA and found to be roughly  consistent with the
experimental data \citep{Rokyta:2005}. This result has been later 
generalised for other fitness distributions \citep{Joyce:2008} and by 
including correlations among fitnesses
\citep{Orr:2006a}. However as the properties of the entire walk are 
required to design a drug or a biomolecule \citep{Bull:2005b} and as 
experimental data on multiple 
adaptive substitutions is becoming 
available \citep{Rokyta:2009,Schoustra:2009}, it is important to
extend the existing theory to address the statistical properties of the
entire walk.

With this aim, we study Gillespie's {\it mutational landscape
  model} on rugged fitness landscapes with many local fitness
  optima. An important difference between 
  our work and the 
previous ones is that here we start the
adaptive walk with low fitness to describe the adaptation process in novel
environments such as when antibiotics are introduced 
\citep{Maclean:2009,Mcdonald:2010} whereas the initial 
fitness is assumed to be high in other studies
\citep{Gillespie:1991,Orr:2002,Orr:2006a,Joyce:2008}. Several numerical
\citep{Gillespie:1991,Orr:2006a} and experimental studies 
 \citep{Rokyta:2009,Schoustra:2009} have indicated that only a few
 steps are  required to  reach a local optimum. In a simple adaptation
 model that assumes the mutational neighborhood to remain unchanged 
 during the entire adaptive walk \citep{Gillespie:1983},  the
  average number of steps to a local fitness peak has been calculated
  analytically for various fitness 
  distributions and shown to increase logarithmically with the rank of
  the initial 
 sequence \citep{Neidhart:2011}. However here we work with a 
 more realistic mutation scheme in which a new suite of mutants is
 created in each adaptive step. For generic fitness distributions, we
 argue that the average number of  adaptive steps increases
  logarithmically with sequence length with a prefactor that depends on the 
 choice of fitness distribution. Although our argument does not
 capture the proportionality constant correctly, the logarithmic
 dependence is seen to be in excellent agreement with the simulation
 results. We also present detailed results on the 
 statistical  properties of entire walk for  
 exponentially and uniformly distributed fitnesses as these two
 distributions  lend themselves to an analytic treatment and are also
 consistent with  the experiments 
 \citep{Eyrewalker:2007,Rokyta:2008}. Following the approach of
 \citet{Flyvbjerg:1992}, we write a recursion relation  
for the fitness distribution of {\it fixed}
beneficial mutations at an adaptive step which is valid for long
sequences and fitness distributions with a finite mean. 
A similar distribution has been calculated in the clonal
 interference regime in which multiple mutants are produced per
 generation \citep{Rozen:2002} while here we work in the weak mutation
 regime.  For the above mentioned distributions, we also find the
 distribution of walk length. The  
 average walk length calculated using this approach gives a prefactor 
 consistent with the  numerical results. 

Although for most of the article we work with uncorrelated
 fitnesses and assume 
 that the distribution of the fitness does not change during the
 course of evolution, the effect of correlations is also discussed. As
 experiments support an intermediate degree of correlations in 
 fitness landscapes  \citep{Carneiro:2010, Miller:2011} and
 changing fitness distributions may be modeled by correlated fitnesses 
 \citep{Orr:2006a}, we calculate the average number of  
 steps to an optimum on a fitness landscape generated by the block
 model of correlated fitnesses in which a sequence 
 is divided into several independent blocks and correlations arise
 when two sequences share some blocks \citep{Perelson:1995}. The
 average walk length has been measured using  numerical 
 simulations in a block model in \citet{Orr:2006a} and it was speculated 
 that the average number of adaptive steps is independent of the
 underlying fitness 
 distribution and increases linearly with the number of blocks. We show 
 that while the latter result is roughly correct, the average number
 of steps to a local optimum is not independent of the fitness
 distribution which is a 
 consequence of the result discussed above for the uncorrelated
 fitness landscapes.

%============================================================================
%Models
%============================================================================

\bigskip
\centerline{MODELS AND METHODS}
\bigskip

{\bf Uncorrelated and correlated fitness landscapes:} An uncorrelated
fitness landscape can be generated by assigning a fitness to a
sequence independent of that of other sequences. The fitnesses are
sampled from a common distribution $p(f)$ with support on the interval
$[ l, u]$.   
Although the full distribution of absolute
fitness is unknown, one can obtain an insight into its nature 
through the distribution  of beneficial mutations which 
has been measured in several theoretical and experimental studies 
\citep{Eyrewalker:2007}. A theoretical argument suggests that since
good mutations are  rare, their distribution is 
governed by the upper tail of the fitness distribution $p(f)$
\citep{Gillespie:1991}. 
It is known from the extreme value theory (EVT) for independent and
identically distributed (i.i.d.) random variables  that 
the asymptotic distribution of the extreme value can be one of 
the following three types \citep{David:2003}: Fr{\'e}chet for
algebraically decaying underlying distributions, Gumbel for unbounded
distributions decaying faster than a power law and Weibull for bounded
distributions. In order to be consistent with this result, 
we make the following choices for the fitness distributions:
\begin{numcases} {p(f)=}
(\delta-1) (1+f)^{-\delta}~&,~$\delta > 2$ ~~~~~~~~~\textrm{(Fr{\'e}chet)} \label{pF} \\
 \gamma f^{\gamma-1} e^{-f^\gamma} ~&,~$\gamma > 0$ ~~~~~~~~~\textrm{(Gumbel)}  \label{pG}\\
\nu (1-f)^{\nu-1} ~&,~$ \nu > 0, f < 1$ ~\textrm{(Weibull)} \label{pW}
\end{numcases}
The condition $\delta > 2$ in (\ref{pF}) is imposed to keep 
the transition rate (\ref{Tc}) finite (as explained later). 
The last two fitness functions (\ref{pG}) and (\ref{pW}) are of
particular interest as several 
experimental results on the distribution of
beneficial mutations have been found to lie in the Gumbel domain
\citep{Imhof:2001,Sanjuan:2004a, 
  Rokyta:2005,Kassen:2006,Maclean:2009} and a recent work finds a best
fit for the distribution of beneficial effects to a uniform
distribution which lies in the Weibull domain \citep{Rokyta:2008}.

We also study adaptive walks on correlated fitness landscapes which
are generated using a block model \citep{Perelson:1995} in which a sequence
of length $L$ is divided into $B$ blocks of 
equal size $L_B=L/B$. The block fitness is an i.i.d. random variable
chosen from the distribution $p(f)$ and the sequence fitness is
obtained on averaging over the fitnesses of the blocks in the
sequence. If two sequences share one or more block, their 
fitnesses are correlated. The correlations can be tuned by changing
the number of blocks: If the number of blocks $B=1$, sequence
fitnesses are completely uncorrelated while $B=L$ gives strongly 
correlated fitnesses. It should be noted that the extreme value
distribution of correlated fitnesses may change from the corresponding 
i.i.d. class 
even if correlations are weak \citep{Jain:2009,Jain:2011b}.   
In the following discussion, we assume that the sequence
  fitnesses are uncorrelated and deal with the correlated fitnesses 
  in the last subsection of this section.

{\bf Adaptive walk model for long sequences:}  
We work with haploid binary sequences of length $L$ in the strong
selection-weak 
mutation (SSWM) regime.  If $N$ is the population size, the SSWM 
regime corresponds to $N s \gg 1, N \mu \ll 1$ where $s$ is the selection
coefficient and $\mu$ is the mutation probability per locus per
generation. Since the 
expected number of mutants produced per generation  is much smaller
than one, mutations occur sequentially and double and higher 
mutations may be neglected. Thus the mutational neighbourhood of a
sequence is limited to $L$ mutants which are single mutation away from
it. If the fitnesses of the wild type sequence and its $L$
one-mutant neighbors are arranged in a 
descending order with the best fitness assigned the rank
$1$, the transition 
probability that the population moves from the wild type with fitness
rank $i$ and value $f_i$ to a mutant with rank $j < i$ and  
value $f_j$ is proportional to the fixation probability 
 which is well approximated by $2 
(f_j-f_i)/f_i$ in the strong selection
limit \citep{Gillespie:1991}. The normalised transition
probability from fitness $f_i$ to fitness $f_j$ is given by 
\be
T(f_j \leftarrow f_i) = \frac{f_j-f_i}{\sum_{k=1}^{i-1} f_k -f_i} ~,~1
\leq  j \leq i-1
\label{Td}
\ee
Once the population has moved to a mutant sequence with fitness $f_j$
with probability $T(f_j \leftarrow f_i)$, it produces a 
set of new mutants which are rank ordered and chosen according to
(\ref{Td}) and the process repeats itself until the
population reaches a local optimum whose nearest 
neighbors are all less fit than itself. 
Note that the parameters $N$ and $\mu$ have dropped out of
the picture and the  properties of the model depend on the sequence
length (or the initial rank)  and the distribution of sequence fitnesses. 

The model described above has been studied using (\ref{Td}) and EVT in
previous works 
\citep{Gillespie:1991,Orr:2002,Orr:2006a,Joyce:2008} 
assuming the initial fitness to be high (small $i$). In contrast, we 
start with a low fitness and write a  
recursion relation for the probability $P_J(f)$ that an adaptive walk has   
at least $J$ steps and the fitness is $f$ at the $J$th step, following 
 \citet{Flyvbjerg:1992} who studied this distribution for
random adaptive walks (see Appendix A). In the following discussion, it is
assumed that the sequence length is large 
which allows the following two simplifications: first, the events in
which a sequence is backtracked can be ignored and second, the
transition rates can be written in 
terms of absolute fitnesses instead of fitness ranks. 
Consider a population at the $J$th adaptive step and with fitness  
$h$. It can proceed to the next step provided 
at least one fitter mutant is available. If $q(h)=\int_l^h  dg~
p(g)$, this event occurs with a
probability $1-q^{L}(h)$ where it is assumed 
that at each step in the evolutionary
process, $L$ novel mutants are available which have not been
encountered before.  While this is true at the 
first step, the number of novel mutants is $L-1$ at the second step
since one of the mutants is the parent sequence itself which is not an
allowed descendant as the walk always proceeds uphill. In fact for any $J
\geq 2$, some of the mutants have already been probed but the error
introduced by ignoring this complication is of the
 order of $1/L$ which is negligible for large $L$ \citep{Flyvbjerg:1992}. 
Then for long sequences we can write 
\bea
P_{J+1}(f) = \int_l^f dh ~p(f) T(f \leftarrow h)~(1-q^{L}(h)) P_J(h)
~,~ J \geq 0 
\label{PJfns1}
\eea
where $p(f) T(f \leftarrow h)$ gives the probability that a 
mutant with fitness $f > h$ is chosen. 
Furthermore for large $L$, it is a good approximation to replace the
sum in the denominator of (\ref{Td}) by an integral and we may write   
\be
T (f \leftarrow h)= \frac{f-h}{\int_h^u dg~(g-h) ~p(g)} ~,~f > h
\label{Tc}
\ee
Thus we work with absolute fitnesses instead of fitness
ranks. 
Since the transition probability (\ref{Tc}) is undefined for
slowly  decaying fitness distributions $p(f) \sim f^{-\delta}~,~
\delta \leq 2$,  we restrict $\delta > 2$ in (\ref{pF}). 
Using (\ref{Tc}) in
(\ref{PJfns1}), we finally obtain
\bea
P_{J+1}(f)= \int_l^f dh~\frac{(f-h) p(f)}{\int_h^u dg~(g-h) ~p(g)}~
 (1-q^{L}(h) ) P_J(h) ~,~ J \geq 0
\label{PJfns}
\eea

Equation (\ref{PJfns}) is the central equation of this article and we
will employ it to obtain various results on the statistical properties
of adaptive walks. In the following, we assume the initial condition
$P_0(f)=\delta(f)$ corresponding to zero initial fitness. As $P_J(f)$ obeys an
integral equation which are harder to analyse, we may try to write a
differential equation for $P_J(f)$. Differentiating (\ref{PJfns})
with respect to $f$, we get: 
\bea
P'_{J+1}(f) &=& \int_l^f dh~\frac{(f-h) p'(f)+p(f)}{\int_h^u dg~(g-h) ~p(g)}~
(1-q^{L}(h)) P_J(h)~,~ J \geq 0 \label{1deriv}  \\
P^{''}_{J+1}(f) &=&
 \int_l^f dh~\frac{(f-h)
  p''(f)+2 p'(f)}{\int_h^u dg~(g-h)~p(g)}  (1-q^{L}(h)) P_J(h) 
 \nonumber \\
&+& \frac{p(f)  (1-q^{L}(f)) }{\int_f^u dg~(g-f) ~p(g)} P_J(f)~,~ J \geq 1
\label{2deriv}
\eea
where prime denotes a $f$-derivative. On using (\ref{PJfns}) and
(\ref{1deriv}) in (\ref{2deriv}), we find
\bea
P^{''}_{J+1}(f) &=& 2 \frac{p'(f)}{p(f)}
P'_{J+1}(f)+ \left[ \frac{p''(f)}{p(f)} -2 \left( \frac{p'(f)}{p(f)}
  \right)^2 \right] P_{J+1}(f)  \nonumber  \\ 
&+& 
\frac{p(f) (1-q^L(f))}{\int_f^u dg~(g-f)~p(g)} P_J(f)~,~ J \geq 1
\label{22deriv}
\eea
The first derivative term in the above equation can be eliminated by
writing $P_J(f)=p(f) {\tilde P}_J(f)$ which finally yields
\be
{\tilde P}^{''}_{J+1}(f)=\frac{p(f) (1-q^L(f))}{\int_f^u
 dg~ (g-f)~p(g)} {\tilde P}_J(f)~,~ J \geq 1
\label{final}
\ee

In this article, we will restrict our attention to exponentially and
uniformly distributed fitnesses as these two fitness distributions are 
consistent with the  available empirical data. 
We show that due to (\ref{final}), a second order ordinary
differential equation is 
obeyed by a generating function of $P_J(f)$ for
these two distributions which can be solved within an approximation 
subject to the following boundary conditions: 
\bea
P_J(f)|_{f=l} &=& 0  ~,~J \geq 1\label{bc1} \\
P'_J(f)|_{f=l} &=& \frac{p(l)}{\int_l^u dg~g ~p(g)}~
\delta_{J,1}
\label{bc2}
\eea
where (\ref{bc1}) is a direct consequence of (\ref{PJfns}) and the
equation (\ref{bc2}) arises on using the initial condition in
(\ref{1deriv}).

Besides $P_J(f)$, we also find the walk length
distribution $Q_J$ and the average fitness ${\bar f}_J$ at the $J$th
step which can be related to $P_J(f)$ as explained below. 
Integrating  over $f$ on both sides of (\ref{PJfns}), we get 
\bea
P_{J+1} &=&\int_l^u df~P_{J+1}(f) \\
&=& \int_l^u dh \int_h^u df~\frac{(f-h) p(f)}{\int_h^u dg (g-h) p(g)}~
 (1-q^{L}(h) ) P_J(h) \\
&=& \int_l^u dh~(1-q^{L}(h)) P_J(h)=P_J-\int_l^u dh~q^{L}(h) P_J(h)
\label{PJns}
\eea
Then the walk length probability $Q_J$ that
exactly $J$ steps are taken is given by
\be
Q_J=P_J-P_{J+1}=\int_l^u dh~q^{L}(h) P_J(h)
\label{QJns}
\ee
with $Q_0=0$ since the initial fitness is zero. 
The above equation has a simple interpretation:  Since $P_J(h)$ is the
probability that at least $J$ steps are taken and the fitness at the
$J$th step is $h$, exactly $J$ steps will 
be taken if all the $L$ mutants of the sequence at the $J$th step carry
a fitness smaller than $h$ from which (\ref{QJns}) follows. The
average walk length ${\bar J}= 
\sum_{J=0}^{2^L} J Q_J \approx \sum_{J=0}^\infty J Q_J$ for large $L$.  
The average fitness ${\bar f}_J$ is defined as ${\bar f}_J
=\int_l^u df~f P_J(f)$.  Using (\ref{PJfns}), we can write 
\bea
{\bar f}_{J+1} &=&\int_l^u df f \int_l^f dh ~\frac{(f-h) p(f)}{\int_h^u dg
  (g-h) p(g)}~ (1-q^{L}(h)) P_J(h)  \\
&=& \int_l^u dh \frac{(1-q^{L}(h)) P_J(h)}{\int_h^u dg
  (g-h) p(g)} \int_h^u df f (f-h) p(f)
\label{FJns}
\eea
Note that neither (\ref{QJns}) nor (\ref{FJns}) are closed equations. 

Our analytical results are also compared with numerical simulations
which were performed using an exact procedure for $L \leq 10$ and an
approximate 
method outlined in \citet{Orr:2002} for larger $L$. We refer the
reader to Appendix B for details.

%============================================================================
%Results
%============================================================================

\bigskip
\centerline{RESULTS}
\bigskip

\begin{figure}[ht]
\begin{center}
\includegraphics[width=0.6 \linewidth,angle=0]{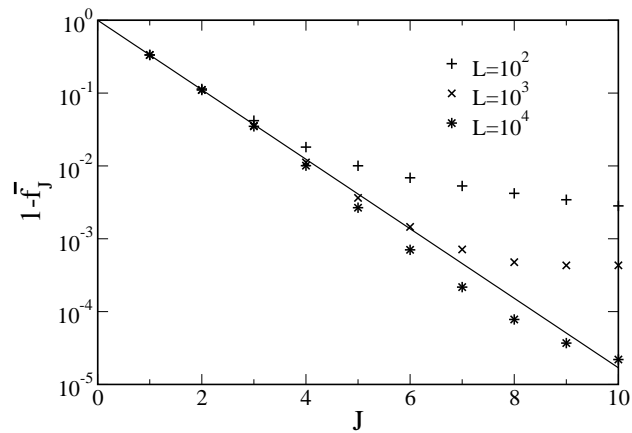}
\caption{Evolution of average fitness with the number of
    adaptive steps starting from zero initial fitness obtained
    numerically (points) and compared with the average fitness in
    infinite sequence 
    length limit (lines) for (a) power law distributed fitness with
    $\delta=6$, equation (\ref{FJF})
    (b) exponentially, equation (\ref{FJexp}) and (c) uniformly
    distributed fitness, equation (\ref{FJuni}).}
\label{FJ}   
\end{center}
\end{figure}

{\bf Average fitness and walk length for general fitness
  distributions:} For a broad class 
of fitness distributions, the average fitness for an infinitely long sequence
can be computed. Although this limit is  biologically
unrealistic, it provides a good 
approximation to the average fitness ${\bar f}_J$ for small $J$ (see
Fig.~\ref{FJ}) as the population can not sense the finiteness of
sequence length far from the local optimum. On taking the limit $L \to
\infty$ in (\ref{FJns}) and denoting the 
average fitness in this limit by $F_J$, we obtain
\be
F_{J+1}=  \int_l^u dh~\frac{\int_h^u df~f (f-h) p(f)}{\int_h^u dg~
  (g-h) p(g)}~P_J(h)|_{L \to \infty} 
\label{FJns2}
\ee

\noindent Algebraically decaying fitness distributions: On
substituting (\ref{pF}) in (\ref{FJns2}) and performing the integrals
involving $p(f)$, we get 
\be
F_{J+1} = \int_0^\infty dh~\frac{2+ (\delta-1) h}{\delta-3}
~P_J(h)|_{L \to \infty}
= \frac{2}{\delta-3}+\frac{\delta-1}{\delta-3} F_J~,~\delta > 3 
\ee
where we have used that $P_J|_{L \to \infty}=1$ due to (\ref{PJns})
and the initial condition $P_0=1$. Repeated iteration with $F_0=0$ yields 
\be
F_J=\left( \frac{\delta-1}{\delta-3} \right)^{J}-1
\label{FJF}
\ee
which increases geometrically with $J$.  This result is compared in
Fig.~\ref{FJ}a with the average fitness for finite sequences 
which shows that 
the number of steps up to which ${\bar f}_J$ and $F_J$ match increases
with $L$.

\noindent Exponential fitness distribution: For fitness distributions
given by (\ref{pG}), the equation for $F_J$ does not close except for
$\gamma=1$.  For $p(f)=e^{-f}$, we get
$F_J=2+F_{J-1}$ which gives 
\be
F_J=2 J
\label{FJexp}
\ee
Fig.~\ref{FJ}b shows that the
rate of increase of fitness ${\bar 
  f}_J$ is slower than a constant at larger $J$'s. 

\noindent Bounded fitness distributions: A calculation similar to
above for $p(f)$ in (\ref{pW}) gives  
\be
F_{J+1} = \frac{2 + \nu F_J}{2+\nu} 
\ee
and therefore 
\be
F_J= 1- \left(\frac{\nu}{2+\nu} \right)^{J}
\label{FJuni}
\ee
For uniformly distributed fitness ($\nu=1$), we find that
$1-F_J=3^{-J}$ in good agreement with the numerical data in
Fig.~\ref{FJ} for small $J$.

%....................................................................

We now give an argument to estimate the average walk length ${\bar J}$
using the above results for the average fitness $F_J$ and the EVT
\citep{Flyvbjerg:1992}. We first note that since $P_J|_{L \to
\infty}=1$ for all $J$, every step in the adaptive walk is definitely
taken for infinitely long  sequences and hence the average walk 
length is expected to diverge with $L$. 
For a sequence of finite length, the adaptive walk 
stops when the population has reached a local
optimum whose fitness is the largest among $L+1$ i.i.d. random 
variables. But since the average number of fitnesses with value 
$\geq f$ is given by  $(L+1) (1-q(f))$, at a local optimum we have
\citep{Sornette:2000}
\be
(L+1) \int_{F_{\bar J}}^u df~p(f)= 1
\label{evt}
\ee
where we have approximated ${\bar f}_{\bar J}$ by $F_{\bar J}$. The above
  equation yields   
\begin{numcases} {F_{\bar J} \approx}
L^\frac{1}{\delta-1}-1  &~
\textrm{(Algebraic)}  \\
 \ln L &~ \textrm{(Exponential)} \label{expevt} \\
1-L^{-\frac{1}{\nu}}  &~ \textrm{(Bounded)}
\label{bdevt}
\end{numcases}
On matching the expected fitness $F_{\bar J}$ with the $F_J$ obtained in
the above discussion for various distributions, we get
\begin{numcases} {{\bar J} \approx}
\frac{1}{\delta-1} \frac{\ln L}{\ln (\frac{\delta-1}{\delta-3})} &~
\textrm{(Algebraic)}  \\
\frac{1}{2} \ln L &~ \textrm{(Exponential)} \\
\frac{1}{\nu} \frac{\ln L}{\ln (\frac{2+\nu}{\nu})} &~ \textrm{(Bounded)}
\end{numcases}
Thus the above argument shows that for large $L$, 
\be
{\bar J}\approx \alpha \ln L 
\label{avgJ2}
\ee
where the prefactor $\alpha$ depends on $p(f)$. We note that $\alpha
_{\textrm{algebraic}} <\alpha_{\textrm{exponential}} <
\alpha_{\textrm{bounded}}$ which implies that smaller  
 number of substitutions occur for fat-tailed fitness distributions than
 the bounded ones. To understand this qualitative trend, consider
   the transition 
  probability for the first step given by $T(f \leftarrow 0) p(f) \sim
  f p(f)$.  At large $f$, this probability is higher for slowly 
  decaying distributions and thus a large
  fitness gain occurs initially.  But as the
  probability to exceed the high fitness achieved at the first step is
  small, the   
walk terminates sooner for broad distributions. 

\begin{figure}[ht]
\begin{center}
\includegraphics[width=0.6 \linewidth,angle=0]{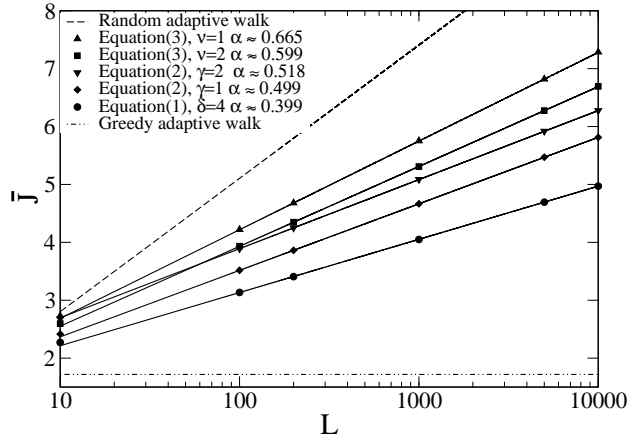}
\caption{Average number ${\bar J}$ of adaptive steps as a function of sequence
  length $L$ for various fitness distributions when the fitnesses are
  uncorrelated. The points show
  the data obtained using numerical simulations and the lines are the
  best fit to the function ${\bar J}= \alpha \ln L+\beta$. The results
  for greedy 
  walk and random adaptive walk (up to an additive constant) are also
  shown. The numerical fit for the prefactor $\alpha$ for 
  exponential and uniform fitness distribution matches well with the
  analytical  
  results given by (\ref{exp_avgJ}) and (\ref{uni_avgJ}) respectively.} 
\label{avg}
\end{center}
\end{figure}

The results of our numerical
simulations for ${\bar J}$ shown in Fig.~\ref{avg} are in agreement
with the logarithmic 
dependence on $L$ but the value of the prefactor does not match  with that
obtained above (except for $p(f)=e^{-f}$).  
The prefactor $\alpha$ is expected to interpolate between the two 
limiting cases of adaptive walks namely greedy walk in which the best
mutant is chosen with probability one and random adaptive walk in
which all better mutants are chosen with equal probability. The former limit is
obtained when $\delta \to 1$ in (\ref{pF}) and the latter when $\nu
\to 0$ in (\ref{pW})  \citep{Joyce:2008}. Since the average walk
length for a greedy walker is a finite constant equal to $e-1 \approx 1.718$
for infinitely long 
sequences \citep{Orr:2003}, the prefactor $\alpha=0$ while $\alpha=1$
for random adaptive walk (see Appendix A). In the following sections, we 
find that $\alpha=1/2$ for exponentially distributed fitness and
$2/3$ for the uniform case which are consistent with the results in
Fig.~\ref{avg} and the analytical results of
  \citet{Neidhart:2011} which are obtained using a simpler version of
  the adaptive 
  walk model considered here.

%-------------------------------------------------------------

{\bf Fitness distribution at the first step for general distributions:}  
If the 
 whole population  is assumed to have an initial fitness $f_0$,  using
 $P_0(f)=\delta(f-f_0)$ in (\ref{PJfns}) we have  
\be
P_{1}(f)= \frac{(f-f_0) p(f) (1-q^{L}(f_0))}{\int_{l}^u dg~g p(g)}~
\propto (f-f_0) p(f)
\label{firststep}
\ee
The above fitness distribution at the first step is nonmonotonic for
all fitness distributions in (\ref{pF})-(\ref{pW}) except for
truncated distributions with $\nu \leq 1$.  The implications of this
result are examined in DISCUSSION.
%........................................................................

{\bf Entire walk with exponentially distributed fitness:} For
$p(f)=e^{-f}$, from (\ref{final}) we obtain
\be
 {\tilde P}^{''}_{J+1}(f)=(1-q^{L}(f)) {\tilde P}_J(f)~,~J \geq 1
\label{exp1}
\ee
where  $q(f)=1-e^{-f}$. 
Due to (\ref{bc1}) and (\ref{bc2}), the boundary conditions are
$P_J(0)=0$ and $P_J'(0)=\delta_{J,1}$.

We define a generating function $G(x,f)= \sum_{J=1}^\infty {\tilde
  P}_J(f) x^J~,~x < 1$  which obeys the following second order ordinary
differential equation: 
\be
G^{''}(x,f)= x (1-q^{L}(f)) G(x,f)
\ee
To arrive at the above equation, we have used that ${\tilde P}_1(f)=f$
which is obtained on using 
  the initial condition in (\ref{PJfns}). 
The generating function $G(x,f)$ obeys a Schr{\"o}dinger equation for
the wave function of a particle in a one-dimensional potential $V(f) \sim
1-q^{L}(f)$ and energy zero \citep{Mathews:1970}. Since $1-q^{L}(f) \approx
  1-e^{-L e^{-f}}$ is 
close to unity for $f \ll \ln L$ and vanishes for  $f \gg \ln L$, the 
potential $V(f)$ decreases smoothly from one to zero and moves
rightwards with increasing $L$. Similar potentials also arise when two
  materials with different transport properties are joined together
  and  in such systems, an  analytical solution is obtained within a
  step function potential approximation
  \citep{Blonder:1982,Schaeybroeck:2009}. We follow this approach 
  here and approximate the distribution $1-q^{L}(f)$  
by the Heaviside theta function $\Theta({\tilde f}-f)$ where ${\tilde
  f}=\ln L$. Within this {\it step distribution approximation}, we
  have
\be
G^{''}(x,f)= 
\begin{cases}
x  G(x,f)  ~&,~f < {\tilde f}\\
0 ~&,~f > {\tilde f}
\end{cases}
\label{expG}
\ee

For $f < {\tilde f}$, the differential equation (\ref{expG}) has a
solution of the form $G_<(x,f)=a_+ e^{\sqrt{x} f}+ a_- e^{-\sqrt{x}
  f}$ which reduces to $G_<(x,f)=c \sinh (\sqrt{x} f)$ since
$G(x,0)=0$ due to $P_J(0)=0$. Since the solution for $f < {\tilde f}$
can not depend on ${\tilde f}$, we appeal to the infinite
sequence length limit to fix the proportionality constant $c$. As
noted earlier,  the distribution
$P_J|_{L \to \infty}=1$ for all $J \geq 0$ which implies that 
\be
\int_0^\infty df~e^{-f} G_<(x,f)= \frac{x}{1-x}
\ee
and therefore 
\be
G_<(x,f)= \sqrt{x} \sinh (\sqrt{x} f)
\label{exp_Gl}
\ee
We check that the boundary condition 
$P_J'(0)={\tilde P}_J'(0)=\delta_{J,1}$ which is equivalent to
$G'(x,0)=x$ is also satisfied by the above solution. 

For $f > {\tilde f}$, the solution $G_>(x,f)=a f+b$ where the
constants of integration $a, b$  can be fixed by matching the  
 solutions $G_<$ and $G_>$ and their first derivative at $f={\tilde
   f}$. Thus the constant $a$ and $b$ are determined by the following
 conditions:
\bea
G_<(x,{\tilde f}) &=& G_>(x,{\tilde f})= a {\tilde f}+b
\label{abconds1} \\
G'_<(x,f)|_{f=\tilde f} &=& G'_>(x,f)|_{f=\tilde f}= a
\label{abconds2}
\eea
A
 simple algebra shows that 
 \be
 G_>(x,f)= x \cosh(\sqrt{x} {\tilde f}) (f-{\tilde f})+ \sqrt{x}
 \sinh(\sqrt{x} {\tilde f})
\label{exp_Gg}
 \ee

\begin{figure}[ht]
\begin{center}
\includegraphics[width=0.6 \linewidth,angle=0]{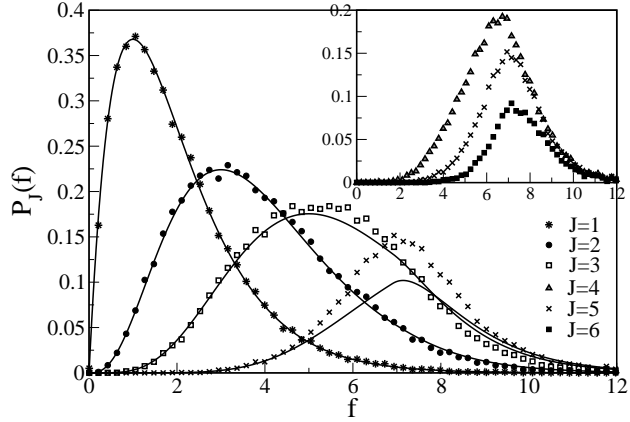}
\caption{Main: Comparison of the distribution $P_J(f)$ for $J=1, 2, 3,
  5$ obtained numerically
  (points) and 
  analytically (lines) given by (\ref{exp_PJf_expr}) for exponentially
  distributed fitness and 
  sequence length $L=1000$. Inset: Numerical data for 
  $P_J(f)$ for $J=4, 5, 6$ to show that the fitness distribution does
  not shift appreciably beyond ${\bar J} \approx 4.6$ as local 
  optimum with average fitness $\approx 7$ is approached.}  
\label{exp_PJf}
\end{center}
\end{figure}

Using the above expressions for $G(x,f)$, the fitness distribution
$P_J(f)$ for the fixed beneficial mutations can be calculated. On expanding
(\ref{exp_Gl}) and (\ref{exp_Gg}) in a power series about $x=0$ and
picking the coefficient of $x^J$, we have 
\be
P_J(f) = \frac{e^{-f} f^{2 J-1}}{(2 J-1)!} \times 
\begin{cases}
 1 &,~ r \leq 1\\
\frac{(2 J-1) r - (2 J-2)}{r^{2 J-1}}&,~ r > 1
\end{cases}
\label{exp_PJf_expr}
\ee 
where $r=f/{\tilde f}$. Figure \ref{exp_PJf} shows our 
numerical results for $P_J(f)$ for the first few adaptive steps. As the
walk proceeds, the distribution 
moves rightwards as expected and its amplitude decreases since the 
probability $q^{L}(f)$ that the walker can not find a better neighbor
approaches unity with increasing $f$.  
Our analytical result 
(\ref{exp_PJf_expr}) is also shown in Fig.~\ref{exp_PJf} for
comparison. 
For $L=10^3$, the step distribution approximation used to find
(\ref{exp_PJf_expr}) gives $1-q^L(f) \approx 1$ for $f < \ln L =6.9$ and
zero otherwise.  However as the probability $1-q^L(f)$
stays close to unity for $f \leq 5$ and decreases gradually to
zero when $f \approx 12$, the distribution  (\ref{exp_PJf_expr}) in
the region $5 < f < 12$ does not match well with the simulation
results but outside this crossover region,  we see a 
good quantitative agreement. We also note that the fitness
distribution does not move appreciably for $J \geq 4$ and is centred
around $f \approx 7$ (see inset of
Fig.~\ref{exp_PJf}). This is because  the average walk length for
$L=10^3$ is about $4.6$ steps (refer  
Fig.~\ref{avg}) and  as the local optimum is
approached, the fitness distribution of fixed beneficial mutation
remains centred close to the typical fitness of the local optimum given by
(\ref{evt}) which is $\ln L \approx 6.9$. This also explains the initial
linear rise in the average fitness followed by a slower increase in
Fig.~\ref{FJ}.

\begin{figure}[ht]
\begin{center}
\includegraphics[width=0.6 \linewidth,angle=0]{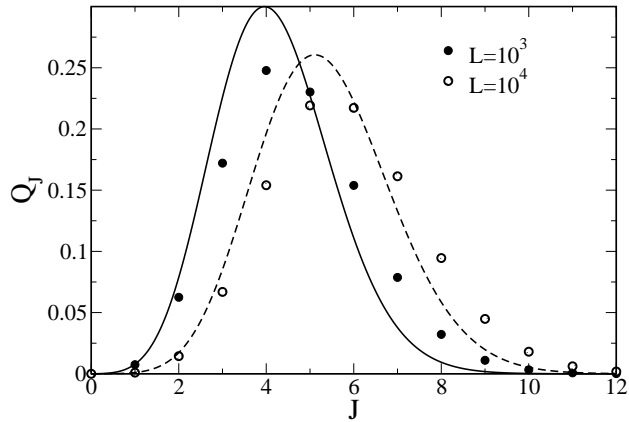}
\caption{Walk length distribution $Q_J$ for $p(f)=
  e^{-f}$ comparing numerical (points) and analytical result (lines)
  given by (\ref{exp_QJ_expr}).} 
\label{exp_QJ}
\end{center}
\end{figure}

We next calculate the walk length distribution $Q_J$ defined by 
(\ref{QJns}). Since $q^L(f)=\Theta(f-{\tilde f})$ within the step 
  distribution approximation discussed above, (\ref{QJns}) reduces  
to 
\be
Q_J=\int_{\tilde f}^\infty df~ P_J(f) 
\ee
On integrating $P_J(f)$
given in (\ref{exp_PJf_expr}), we get 
\be
Q_J= e^{-\ln L} \left[ \frac{(\ln L)^{2 J-2}}{(2 J-2)!}+\frac{(\ln
 L)^{2 J-1}}{(2 J-1)!}\right] ~,~J > 0
\label{exp_QJ_expr}
\ee
This expression is compared with numerical results in
Fig.~\ref{exp_QJ} and shows a reasonable agreement. The average number
of adaptive steps calculated using (\ref{exp_QJ_expr}) is given by  
\be
{\bar J} = \sum_{J=1}^\infty J Q_J \approx \frac{1}{2} \ln L 
\label{exp_avgJ}
\ee
which is in good agreement with the simulation result in
Fig.~\ref{avg}. The width of the distribution $Q_J$ measured using the
variance $\sigma^2=\bar{{J^2}}-{\bar J}^2 \approx \ln L/4$ also increases
with $L$.

%-------------------------------------------------------------

\begin{figure}[ht]
\begin{center}
\includegraphics[width=0.6 \linewidth,angle=270]{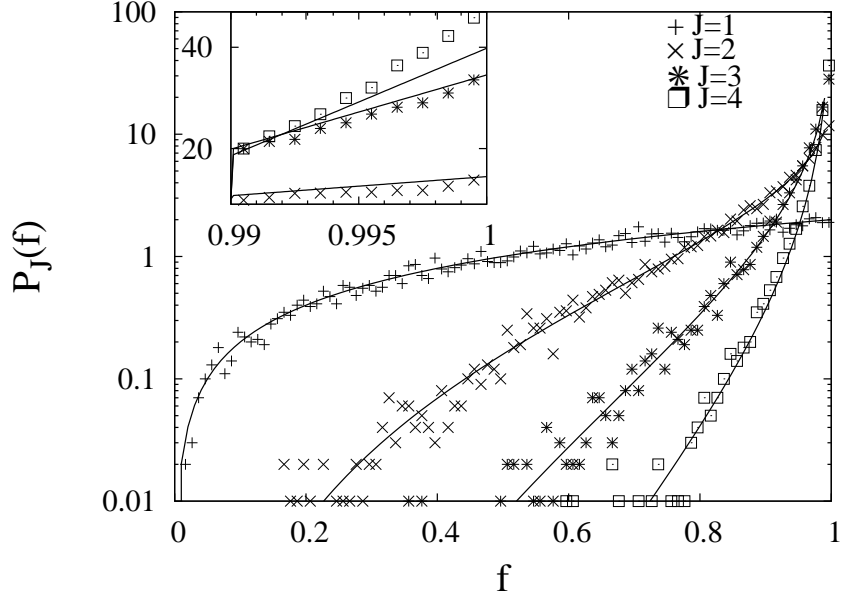}
\caption{Comparison of the distribution $P_J(f)$ for $J=1, 2, 3,
  4$ obtained numerically
  (points) and 
  analytically (lines) given by (\ref{uni1})-(\ref{uni4}) for uniformly
  distributed fitness and 
  sequence length $L=100$. The distribution for $f
  \leq {\tilde f}$ is shown in the main plot and for $f > {\tilde f}$ in
  the inset.}
\label{uni_PJf2}
\end{center}
\end{figure}

{\bf Entire walk with uniformly distributed fitness:}  
%Since all the derivatives of $p(f)$ vanish 
For $p(f)=1$, since $P_J(f)={\tilde P}_J(f)$, the differential
  equation (\ref{final}) reduces to 
\bea
P^{''}_{J+1}(f)= \frac{1-f^L}{\int_f^1 dg~(g-f)}~
P_J(f)= \frac{2 (1-f^L)}{(1-f)^2} P_J(f) ~,~J \geq 1
\eea
with boundary conditions $P_J(0)=0$ and $P_J'(0)=2 \delta_{J,1}$.
As before, we define a generating  
function $G(x,f)= \sum_{J=2}^\infty x^{J-2} P_J(f)$ which obeys the
following second order ordinary differential equation: 
\be
G^{''}(x,f)= \frac{2 (1-f^L)}{(1-f)^2} ~(x G (x,f)+2 f) 
\ee
where we have used that $P_1(f)=2 f$. 
We treat this case also within the step distribution approximation discussed
earlier. Since the probability $1-f^{L} \approx 1-e^{-L (1-f)}$, we
approximate it by a step function $\Theta({\tilde f}-f)$ where
${\tilde f}=(L-1)/L$. 
For $f < {\tilde f}$, we obtain an inhomogeneous  second order
ordinary differential equation with variable coefficients: 
\bea
G_<^{''}(x,f) &=& \frac{2 x}{(1-f)^2} G_<(x,f)+ \frac{4 f}{(1-f)^2}
\label{uniode}
\eea
This equation can be solved by standard methods (as detailed in Appendix
C) to yield 
\be
G_<(x,f)= a_+ (1-f)^{\alpha_+}+ a_- (1-f)^{\alpha_-}+ u_+(f) (1-f)^{\alpha_+}+
u_-(f) (1-f)^{\alpha_-} 
\ee
where the exponents 
\be
\alpha_{\pm}= \frac{1 \pm \sqrt{1+ 8 x}}{2}
\label{alphapm}
\ee 
The first two terms on the right hand side give the solution of
the homogeneous equation and
the last two terms are the particular integral involving the
variational parameters $u_{\pm}(f)$ given in Appendix C. The constants
of integration 
$a_\pm$ can be obtained using the boundary conditions $G(x,0)=0$ and
$\int_0^1 df~G_<(x, f)= (1-x)^{-1}$. After some straightforward
algebra, we find that
\be
G_<(x,f)= \frac{-2}{x} \left[ \frac{(1-f)^{\alpha_+}
    -(1-f)^{\alpha_-}}{\alpha_+ -\alpha_-}+f \right]
\label{uniGl}
\ee
We verify that the condition $P_J'(0)=0$ for $J > 1$ which
  amounts to $G'(x,0)=0$ is also satisfied.
For $f > {\tilde f}$,  as $G^{''}_>(x,f)=0$, the solution
$G_>(x,f)=a f+ b$ where $a,b$ can be determined using (\ref{abconds1})
and (\ref{abconds2}) to give 
\bea
G_>(x,f) &=&\frac{-2}{x} \left[ \frac{\alpha_- (1-{\tilde f})^{\alpha_- - 1}
    -\alpha_+ (1-{\tilde f})^{\alpha_+ -1}}{\alpha_+ -\alpha_-} + 1\right]
f  \nonumber \\
&-& \frac{2}{x} \left[\frac{(1-{\tilde f})^{\alpha_+}
    -(1-{\tilde f})^{\alpha_-}-\alpha_- {\tilde f} (1-{\tilde
      f})^{\alpha_- - 1} 
    +\alpha_+ {\tilde f} (1-{\tilde f})^{\alpha_+ -1}} {\alpha_+
    -\alpha_-} \right]  
\label{uniGg}
\eea
Explicit expressions for $P_J(f)$ for first few adaptive steps are
given in Appendix C and a comparison between the analytical and
the simulation results is shown in Fig.~\ref{uni_PJf2}.

\begin{figure}[ht]
\begin{center}
\includegraphics[width=0.6 \linewidth,angle=0]{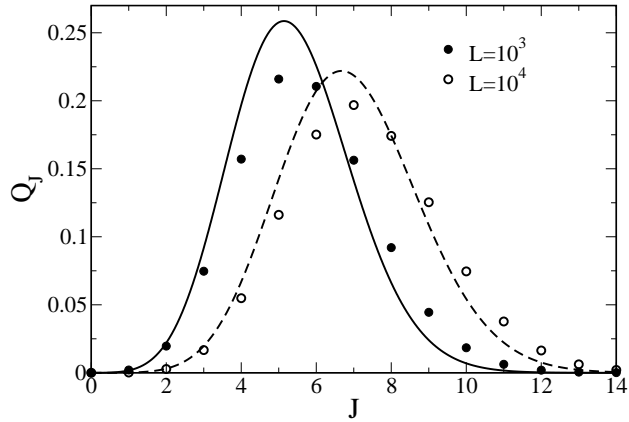}
\caption{Walk length distribution $Q_J$ for uniformly distributed
  fitnesses comparing simulation (points) and analytical result 
  (lines) in (\ref{QJuni_expr}).}
\label{uni_QJ}
\end{center}
\end{figure}

To find the walk length distribution $Q_J=\int_{{\tilde f}}^1 df~
P_J(f)$, we define  
\bea
H(x) &=& \sum_{J=1}^\infty x^J Q_J =x Q_1+ x^2 \int_{\tilde f}^1 df~G_>(x,f) \\
&=& \frac{ x (1-{\tilde f})}{\alpha_- - \alpha_+} \left[(2-\alpha_+) (1-{\tilde
    f})^{\alpha_+} - (2-\alpha_-)(1-{\tilde f})^{\alpha_-}
  \right]
\label{Hx}
\eea
As an explicit expression for $Q_J$ is rather unwieldy, its derivation and 
the expression itself are given in Appendix C and a comparison
  with the simulations is shown in Fig.~\ref{uni_QJ}. The average number of
steps is given by  
\be
{\bar J}=  \frac{d H(x)}{dx}\bigg|_{x=1}=\frac{-6 \ln (1-{\tilde f})}{9}
\label{uni_avgJ}
\ee
which shows that for large $L$, the number of adaptive steps grows as
$(2/3) \ln L$ in agreement with the numerical results shown in
Fig.~\ref{avg}. The higher moments can also be found straightforwardly
and we find that the variance ${\bar {J^2}} - {\bar
  J}^2 \approx (10/27) \ln L$ and the skewness of the distribution
decays slowly as
$(\ln L)^{-1/2}$. 

%-------------------------------------------------------------

{\bf Effect of correlations on the number of adaptive steps:} 
We now turn to a discussion of adaptive walk properties when 
the fitnesses are correlated and given by a block model. We compute
the average number ${\bar J}_B(L)$ of adaptive steps given 
by $\sum_{J=1}^\infty J Q_J(L,B)$ where $Q_J(L,B)$ is the probability
that exactly $J$ adaptive mutations occur when a sequence of length
$L$ is divided in $B$ blocks. 

Consider
the distribution ${\cal Q}(m_1,...,m_B)$ which gives the 
joint probability that the $i$th block of length $L_B$ in a sequence
of length $L$ carries $m_i$ adaptive 
mutations where $i=1,...,B$.  
An important property of the block model is 
that this joint distribution factorises, that is \citep{Perelson:1995} 
\be
{\cal Q}(m_1,...,m_B) = \prod_{b=1}^B Q_{m_b}(L_B,1)
\label{prod}
\ee
where $Q_J(L_B,1) \equiv Q_J(L_B)$ is the walk length probability when
the fitnesses are uncorrelated and the sequence length is $L_B$. The
above equation expresses the fact that the block 
fitnesses evolve independently. As only one mutation occurs in the
sequence at any step  
so that all but one block sequence remains unchanged and since the block
fitnesses are i.i.d. random variables, (\ref{prod}) holds. 

\begin{figure}[ht]
\begin{center}
\includegraphics[width=0.6 \linewidth,angle=0]{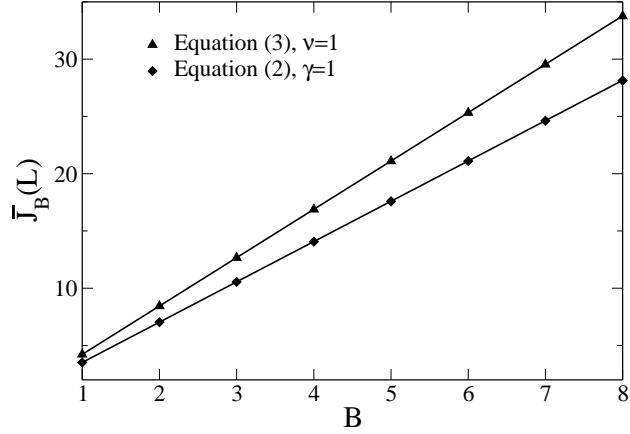}
\caption{Average number ${\bar J}_B$ of adaptive steps as a
    function of block number $B$ for fixed $L/B=100$. The numerical data
    is in excellent agreement with 
    (\ref{linear}) shown by solid line.}
\label{blockfig}
\end{center}
\end{figure}

Since the distribution $Q_J(L,B)$ is given by 
\be
Q_J(L, B)= \sum_{m_1,...,m_B=0}^J {\cal Q}(m_1,...,m_B)
\delta(m_1+...+m_B -J)
\label{prod2}
\ee
it follows that 
\bea
{\bar J}_B (L) &=& \sum_{J=1}^\infty J \sum_{m_B=0}^J Q_{m_B}(L_B)
\sum_{m_1,...,m_{B-1}=0}^{J-m_B} \prod_{b=1}^{B-1} Q_{m_b}(L_B)
\delta(\sum_{b=1}^{B-1} m_b -(J-m_B))  
\nonumber \\
&=&\sum_{J=1}^\infty J \sum_{m_B=0}^J  Q_{m_B}(L_B)
Q_{J-m_B}(L-L_B,B-1) \nonumber
\\
&=& \sum_{m=0}^\infty Q_m(L-L_B,B-1) \sum_{n=0}^\infty (n+m) Q_n (L_B)
\nonumber\\
&=& {\bar J}(L_B)  + \sum_{m=1}^\infty m
Q_m(L-L_B,B-1) \nonumber \\
&=& {\bar J}(L_B) + {\bar J}_{B-1}(L-L_B) \nonumber \\
&=& B {\bar J}(L_B)
\label{linear}
\eea
where we have used that $\sum_{J=0}^\infty
Q_J(L,B)=1$ and ${\bar J}$ is the average number
of steps in the adaptive walk for uncorrelated fitnesses. Figure
\ref{blockfig} shows the results of our numerical simulations for
average walk length when the block length $L_B=L/B$ is kept fixed and
the block fitnesses are exponentially and uniformly distributed. For
fixed $L_B$, (\ref{linear}) predicts that ${\bar J}_B$ increases
linearly with $B$ which is in excellent agreement with the 
numerical data.

For large 
$L$, due to (\ref{avgJ2}) we have 
\be
{\bar J}_B (L) \approx \alpha B \ln(L/B) 
\ee
For small $B$, a linear rise in the average number of steps with the
number of blocks has been seen numerically for exponential-like
distributions and it was inferred that the mean walk length is
independent of underlying fitness distributions \citep{Orr:2006a}. However as
discussed in the previous sections, the average number 
${\bar J}$ depends on the fitness 
distribution $p(f)$ and therefore the average ${\bar J}_B$ is also
nonuniversal.

%============================================================================
%DISCUSSION
%============================================================================

\bigskip
\centerline{DISCUSSION}
\bigskip

In the last few years, several analytical results have been obtained
for the mutational landscape model \citep{Gillespie:1991}. However
many of these results deal 
with the first step in the adaptation process \citep{Orr:2002,
  Orr:2006a, Joyce:2008} and an extension of the
theory to full adaptive walk is necessary. Previous studies also
assume that the process of adaptation starts from a highly fit
sequence which is not 
applicable to situations in which the population is subjected to high
stress and hence has a very low initial fitness
\citep{Maclean:2009,Mcdonald:2010}. In this article, we have obtained 
results for the entire adaptive walk starting from a low initial
fitness but as discussed below, we expect some of these results to
hold for moderately high initial fitness also.

{\bf Walk length distribution and average walk length:}  In previous
works, the walk length distribution for the greedy walk  and the random 
adaptive walk have been studied and found to be  universal in that they are
independent of the underlying fitness distribution $p(f)$. The origin of
this universality property is clear in the light of 
the results of \citet{Joyce:2008} who pointed out that these two
models can be obtained as a limit of (\ref{Td}) which defines the
mutational landscape model.  For the random
adaptive walk, the distribution $Q_J$ for infinitely long sequence vanishes 
(see (\ref{QJ_rw})) and the average walk length diverges 
with sequence length. In contrast, for greedy walk, the walk length
distribution in the $L \to \infty$ limit decreases exponentially fast with 
$J$ for the greedy walk as a result of which 
the average number of steps turns out to be a constant
\citep{Orr:2003,Rosenberg:2005}.

In this article, we have calculated the walk length distribution for
exponentially and uniformly distributed fitnesses and found the
average walk length for general fitness distributions. 
An important conclusion of our study
is that the average number of adaptive steps increases logarithmically with
the sequence length with a prefactor smaller than unity if the walk starts
from zero fitness. Our simulations (not shown) also indicate that if
the initial rank is of order $L$, the average number of steps
increases logarithmically with the rank and with the same
proportionality constant as that for the zero initial fitness
case. Thus for a wild type sequence with initial rank (or 
$L$) of the order $100$, the number of substitutions are expected to
be less than $5$. Although short adaptive walks have been observed in
experiments \citep{Rokyta:2009,Schoustra:2009}, 
more detailed experimental studies testing the
logarithmic dependence would be desirable.  
Although a test of the $L$-dependence of the average walk length may
not be experimentally viable, it should be
possible to study the average walk length as a function of the initial
rank. 

Besides the sequence length, the number of steps to a local optimum
depend on the underlying fitness  
distribution and the fitness correlations also. 
If the fitnesses are uncorrelated, as the numerical data in
Fig.~\ref{avg} shows, the prefactor $\alpha$ in (\ref{avgJ2}) depends
on the shape of the 
fitness distribution and therefore a rather detailed knowledge of the
full fitness distribution (how fast it decays) is required to test
this which is presently unavailable. However one can discern a trend in the
value of $\alpha$: it decreases as the fitness distribution
broadens. This suggests that systems with fitness
distribution in the Gumbel class \citep{Imhof:2001,Sanjuan:2004a,
  Rokyta:2005,Kassen:2006,Maclean:2009} will register shorter
walks than those  in the Weibull domain \citep{Rokyta:2008}. 
As shown here in the block model of correlated fitnesses, the average
number of adaptive steps increases as the  
number of blocks (and hence fitness correlations) increase. This is in
accordance with the expectation that on a smooth correlated fitness
landscape, as the local optima are less common \citep{Perelson:1995},
there is a less chance to get trapped and therefore uphill walk can
last longer \citep{Weinberger:1991,Kauffman:1993,Orr:2006a}.

{\bf Distribution of fixed beneficial mutations during the walk:} 
The fitness distribution $P_J(f)$ has not been studied in previous
theoretical studies of adaptive walks in the SSWM limit and here we have
computed this fitness distribution analytically using the recursion
relation (\ref{PJfns}). The fitness distribution 
at the first step given by (\ref{firststep}) can give a qualitative
idea  about the shape of $p(f)$.  For most fitness distributions, $P_1(f)$ is 
expected to be nonmonotonic but for bounded distributions which
diverge at the upper limit or the uniform distribution,  $P_1(f)$
increases monotonically towards the upper bound.  
An inspection of the experimental data  of 
\citet{Rokyta:2005} shows the fitness distribution
at the first step to be nonmonotonic which is consistent with their 
assumption of exponentially decreasing distribution of beneficial
effects. It would be interesting to check if the distribution $P_1(f)$ in
\citet{Rokyta:2008} is monotonic as the data in this study is
consistent with a uniformly distributed fitness.  The above behavior
of $P_1(f)$ is expected to be robust 
in the presence of correlations as at the first step in evolution, the
population has not sensed the correlations in the fitness landscape
\citep{Orr:2006a}. 

For the fitness distribution for the entire walk, we presented an 
analysis for two distributions namely exponential and uniform which are 
consistent with the available experimental data. The distribution
$P_J(f)$ is obtained within a step distribution  
approximation which captures the shape of the fitness 
distribution correctly for the first few steps 
and leads to an accurate estimate of the number of average steps. Our
approximation consists of replacing the 
probability  $1-q^L(f)$ by a step function $\Theta({\tilde f}-f)$ where
${\tilde f}$ is given by (\ref{expevt}) for
exponentially and (\ref{bdevt}) for uniformly distributed
fitnesses. 
For $f \ll {\tilde f}$ and $f \gg   {\tilde f}$, our
approximate solution matches the simulation results well for any
$J$. With increasing $J$, the distribution $P_J(f)$ shifts towards
higher fitnesses and peaks about ${\tilde f}$ for larger
$J$'s. As explained earlier, the fitness ${\tilde f}$ is reached when
$J$ is close  
to ${\bar J} \propto \ln L$ and therefore we expect our approximation
to work well for $J \ll \ln L$.

When the underlying fitness distribution is exponential, we find that the 
 fitness distribution of the fixed beneficial mutation also has an 
 exponential tail (see (\ref{exp_PJf_expr})). The robustness of this
 result {\it i.e.} whether any fitness distribution in the Gumbel class
 exhibits exponential tail for $P_J(f)$ is however not clear. 
 For uniformly distributed fitnesses, as the width of the distribution 
$1-q^L(f)$ decreases with increasing $L$, the step  
 distribution  approximation works better in this case than in 
 the exponential case where the width is a constant (compare
 Figs.~\ref{exp_QJ} and \ref{uni_QJ}). 
The properties of multiple steps in an adaptive walk have been measured
in some recent experiments \citep{Rokyta:2009,Schoustra:2009} and a
detailed analysis of the experimental results would be very welcome.  
On the theoretical front, an extension of the results described above
to distributions other than uniform and exponential would be
desirable. We have recently made some progress in this direction
 and the results will appear elsewhere.

\begin{figure}[ht]
\begin{center}
\includegraphics[width=0.6 \linewidth,angle=0]{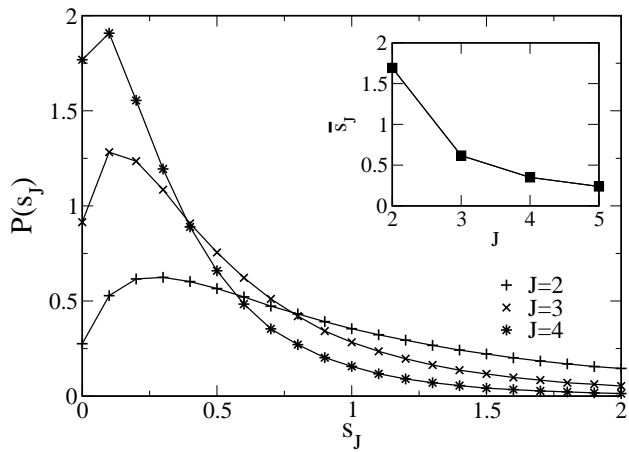}
\caption{Distribution $P(s_J)$ of selection coefficient $s_J$ for
    $L=1000$ and $p(f)=e^{-f}$. The inset shows the decay in average
    selection coefficient ${\bar s}_J$ as a function of $J$. The points are
    joined by line to guide the eye. }
\label{PJs}
\end{center}
\end{figure}

Another interesting question concerns the distribution $P(s_J)$
  of the selection coefficient $s_{J}= (f_{J}-f_{J-1})/f_{J-1}$ at the
  $J$th step in the adaptive walk. As we start with zero fitness, the
  selection coefficient is defined for $J \geq 2$. Our preliminary
  numerical results for $P(s_J)$ are shown in Fig.~\ref{PJs} for the first
  few steps in the walk and we observe that the typical selection
  coefficient decreases as the walk proceeds. This
  behavior matches qualitatively with the experimental results of
  \citet{Schoustra:2009}. A theoretical analysis of the distribution
  $P(s_J)$ requires the joint distribution of the fitness at step
  $J-1$ and $J$ and we hope to address this question in a future work.

Acknowledgement: K.J. thanks J. R. David for helpful suggestions,
J. Krug for comments on an earlier version of the manuscript and
useful correspondence and KITP, Santa Barbara  for 
hospitality  and support under the NSF Grant No. PHY05-51164.  The
authors also thank L. Wahl for suggestions to improve the manuscript.

%============================================================================
%APPENDIX
%============================================================================
\bigskip
\centerline{APPENDIX A: RANDOM ADAPTIVE WALK}
\label{AppA}
\bigskip

In this Appendix, we briefly review the known results for random adaptive
walk  in which all better mutants are chosen with
equal probability
\citep{Macken:1989,Flyvbjerg:1992,Kauffman:1993}.  The 
probability distribution $P_J(f)$ obeys the
following recursion relation \citep{Flyvbjerg:1992}:
\be
P_{J+1}(f)= \int_l^f dh~\frac{p(f)}{\int_h^u dg~ p(g)}~
\left[ 1-q^{L}(h) \right] P_J(h)
\label{PJ_rw}
\ee
where $q(f)=\int_l^f dg~p(g)$. 
A change of variable from the fitness
$f$ to the cumulative probability $q(f)$ gives 
\be
{  P}_{J+1}(q)=\int_0^q dq'\frac{1-q'^L}{1-q'}~
{  P}_J(q')
\label{PJ2_rw}
\ee
Since the walk length distribution for the random adaptive walk also
obeys (\ref{QJns}), we have  
\bea
Q_J=\int_l^u dh~q^{L}(h) P_J(h)=\int_l^u dq~q^{L} {  P}_J(q)
\label{QJ_rw}
\eea
which shows that $Q_J$ is a {\it universal distribution} in that it is
independent of 
the underlying fitness distribution $p(f)$. Note that for infinitely
long sequences, the probability $Q_J=0$ as in the mutational landscape
model. Differentiating (\ref{PJ2_rw}) with respect to $q$ immediately gives 
\be
\frac{ d {  P}_{J+1}(q)}{dq}=\frac{1-q^{L}}{1-q}~{  P}_J(q)=
\sum_{n=0}^{L} q^n ~{  P}_J(q)
\ee
The generating function $G(x, q)=\sum_{J=1}^\infty x^J {  P}_J( q)$
then obeys the following {\it first order} differential equation: 
\be
G'(x, q)- x {  P}_1'( q)= x \frac{1-q^{L}}{1-q} G(x, q)
\ee
For the initial condition $P_0(f)=\delta(f)$, we have $P_1( q)=1$ and
due to (\ref{PJ2_rw}), the distribution ${P}_J(0)=0$. Solving the
above  differential equation using these boundary conditions gives 
$G(x, q)=x e^{x H_L( q)}$ where $H_L( q)=\sum_{k=1}^L q^k/k$ and hence
the distribution ${ 
  P}_J( q)$ is given by \citep{Flyvbjerg:1992} 
\be
{  P}_J( q)= \frac{H^{J-1}_L( q)}{(J-1)!}
\ee
Since the product  $q^{L} {  P}_J(q)$ in (\ref{QJ_rw}) peaks around
$q=1$, using $H_L( q) \approx \ln L$ for $q$ close to unity for finite but
long sequences and performing the integral in (\ref{QJ_rw}), we get
\be
Q_J \approx e^{-\bar J} \frac{{\bar J}^{J-1}}{(J-1)!}
\label{QJ_rw2}
\ee
where ${\bar J}=\ln L$. Thus the walk length distribution is a Poisson
distribution 
(in $J$) with mean ${\bar J}=\ln L$ \citep{Flyvbjerg:1992}.

\bigskip
\centerline{APPENDIX B: SIMULATION PROCEDURE}
\bigskip

For short sequences of length $L \leq 10$ and uncorrelated fitnesses,
a randomly chosen sequence was assigned a fitness equal to zero. Then
the rest of the fitness landscape comprising of $2^L-1$ fitnesses was 
generated by drawing random variables independently from a common
distribution $p(f)$. The transition probability from the initial
sequence to each of the better sequences  
among the $L$ nearest neighbors was calculated according to (\ref{Td})
and the fixed sequence at the first step in the adaptive walk was
chosen. Then the transition probability 
from the chosen mutant  
sequence to its better neighbors was calculated and this process was
repeated until a fitter sequence was not available. 

To simulate sequences with length  $L \gtrsim
10^2$, we followed an approximate procedure outlined in
\citet{Orr:2002} as the total number of sequences $2^L$ is
prohibitively large for long sequences. Starting with zero fitness, 
$L$ i.i.d. random variables were generated and a higher 
fitness $f$ was chosen according to the transition probability 
(\ref{Td}). During the next step in the process, $L$ new
i.i.d. random variables were 
generated and the transition probability from $f$ to a better fitness
was calculated. These steps were repeated until the new set of random
fitnesses does not exceed the currently fixed fitness. 
The block model was 
simulated to generate weakly correlated fitnesses by assigning
independent fitnesses to each block sequence. In all the simulations,
the data was collected using $10^6$ 
independent realisations of the fitness landscape.

\bigskip
\centerline{APPENDIX C: DERIVATIONS FOR UNIFORMLY DISTRIBUTED FITNESS}
\bigskip

{\bf Solution of differential equation  \ref{uniode}:}
The generating function $G_<(x,f)$ obeys the following inhomogeneous
second order differential equation:
\be
G^{''}(x,f)-\frac{2 x}{(1-f)^2} G(x,f)= \frac{4 f}{(1-f)^2}
\ee
where we have dropped the subscript for brevity. The general solution
of such differential equations 
is a linear combination of the general solution $G_H(x,f)$ of the
homogeneous equation obtained by setting the right hand side equal to
zero and the particular solution $G_P$ 
of the inhomogeneous equation \citep{Mathews:1970}. The homogeneous
solution is of the form 
\be
G_H(x,f)=a_+(1-f)^{\alpha_+}+ a_- (1-f)^{\alpha_-}  
\ee
where $\alpha_\pm$ are the
solutions of the quadratic equation $\alpha^2-\alpha-2 x=0$ and given by
(\ref{alphapm}). The particular solution is found using the method of
variation of parameters and is of the form $G_P(x,f)= u_+(x) (1-f)^{\alpha_+} +
u_-(x) (1-f)^{\alpha_-}$ where the functions $u_\pm (f)$ obey the
following first order differential equations \citep{Mathews:1970}:
\bea
u'_+(f) (1-f)^{\alpha_+}+u'_-(f) (1-f)^{\alpha_-} &= &0 \\
\alpha_+ u'_+(f)  (1-f)^{\alpha_+ -1}+\alpha_- u'_-(f) (1-f)^{\alpha_-
    -1} &=& \frac{4f}{(1-f)^2}
\eea
On solving the above equations, we obtain
\bea
G_P(x, f) = \frac{4}{\alpha_+ \alpha_-} - \frac{4 (1-f)}{(1-\alpha_+)
  (1-\alpha_-)} =\frac{-2 f}{x}
\eea
Finally using the boundary conditions in the general solution $G_<(x,f)=
G_P(x,f)+G_H(x,f)$, the desired result (\ref{uniGl}) is obtained.

{\bf Distribution of fixed beneficial mutations:} The fitness
  distribution found using
  (\ref{uniGl}) and (\ref{uniGg}) is given below for the first few
  adaptive steps:
\begin{eqnarray}
P_1(f) &=& 2 f  ~~~~ ,~ f \leq 1 \label{uni1} \\
P_2(f) &=& \left \{ \begin{array}{ll}
-8 f + 4 (f-2) \ln(1 - f)  & ,~f \leq {\tilde f} \\
\frac{4 {\tilde f} (f + {\tilde f}-2)}{1 - {\tilde f}} + 4 (f-2)
\ln (1 - {\tilde f}) & ~,~f > {\tilde f}
\end{array} \right. \\
P_3(f) &=& 4 \left \{ \begin{array}{ll}
12 f + \ln(1 - f) (12 - 6 f + f \ln(1 - f)) &,~f \leq {\tilde f} \\
\frac{1}{1-{\tilde f}} \left[6 {\tilde f} (2 - f - {\tilde f}) +
2 (6 - (6 - {\tilde f}) {\tilde f} - f (3 - 2 {\tilde f})) \ln(1 -
  {\tilde f}) \right. \\
+ \left. f (1 - {\tilde f}) \ln^2(1 - {\tilde f}) \right] &,~f > {\tilde f}
\end{array} \right. \\
P_4(f) &=& \frac{-8}{3} \left \{ \begin{array}{ll}
120 f +60 (2 - f) \ln(1 - f)+12 f \ln^2(1 - f) +
(2 - f) \ln^3(1 - f)  ~,~f \leq {\tilde f} \\
\frac{1}{(1-{\tilde f})} \left[60 {\tilde f} (2 - f - {\tilde f}) -
 12  (f (5 - 3 {\tilde f}) - 2 (5 - (5 -
     {\tilde f}) {\tilde f})) \ln(1 - {\tilde f})\right.  \\ 
+  \left.  3 (f (2 - 3 {\tilde f}) +
        (2 - {\tilde f}) {\tilde f}) \ln^2(1 - {\tilde f})+  (2 -
	f) (1 - {\tilde f}) \ln^3(1 - {\tilde f})\right]~,~f >
     {\tilde f}
\label{uni4}
\end{array} \right.
\end{eqnarray}

{\bf Walk length distribution:}  On matching powers
of $x^J$ on both sides in (\ref{Hx}), we get 
\bea
Q_1 &=& e^{-2 \ell} (-1+ 2 e^\ell) \\
Q_2 &=& 2 e^{-2 \ell} (3 + \ell+ (-3 + 2 \ell) e^\ell) \\
Q_3 &=& e^{-2 \ell} \left[-2 (18+ 8 \ell+ \ell^2)+4 e^\ell (9- 5 \ell+
  \ell^2) \right] \\ 
Q_4 &=& \frac{4 e^{-2 \ell}}{3} \left[180+ 84 \ell+ 15 \ell^2+ \ell^3+
  e^\ell (-180+ 96 \ell -21 \ell^2 + 2 \ell^3) \right]
\eea
where $\ell=\ln L$. A general solution of $Q_J$ by this method does
not seem possible but an approximate analytic expression for $Q_J$ can be
obtained as explained below.

From the definition of the generating function $H(x)$ in (\ref{Hx}), it
follows that  
\be
Q_J= \frac{1}{J!} \frac{d^J H(x)}{dx^J} \bigg|_{x=0}
\label{QJ1}
\ee
By the residue theorem for complex variables, we have \citep{Mathews:1970} 
\be
\frac{1}{2 \pi i} \int_C  dz~f(z)= \frac{1}{n!}
\frac{d^n}{dz^n} \left( (z-z_0)^{n+1} f(z) \right) \bigg|_{z=z_0}
\label{res}
\ee
where $z_0$ is a pole of order $n+1$ of the function $f(z)$ and the contour $C$
encloses the singularities of $f(z)$. From (\ref{QJ1}) and
(\ref{res}), we can write 
\be
Q_J= \frac{1}{2 \pi i}\int_C dz~\frac{H(z)}{z^{J+1}}=  \frac{1}{2 \pi
  i} \int_C dz~e^{K(z)} 
\ee
where $K(z)=\ln H(z)-(J+1) \ln z$. We solve this integral by the
method of steepest descent which for large $J$ gives
\citep{Mathews:1970} 
\be
Q_J \approx  \sqrt{\frac{1}{2 \pi K''(z_s) }}
e^{K(z_s)} = \sqrt{\frac{1}{2 \pi K''(z_s) }}
\frac{H(z_s)}{z_s^{J+1}} 
\ee
where prime refers to derivative with respect to $z$. 
In the above equation, $z_s$ is a solution of the equation
\be
\frac{H'(z_s)}{H(z_s)}=\frac{J}{z_s}
\label{zs1}
\ee
and 
\bea
K''(z_s) &=& \left(\frac{H'(z)}{H(z)}
\right)'\bigg|_{z=z_s}+\frac{J}{z_s^2} \\
&=& \left(\frac{H'(z)}{H(z)}
\right)'\bigg|_{z=z_s}+ \frac{1}{z_s} \frac{H'(z_s)}{H(z_s)}
\label{fluct1}
\eea
where prime denotes a derivative with respect to $z$. Since $\alpha_+
> 0$, neglecting the exponentially small term in $(1-{\tilde
  f})^{\alpha_+}$ in (\ref{Hx}), we get
\be
H(z) \approx \frac{e^{-3 \ell/2} e^{\ell y/2} (3 + y) (y^2-1)}{16 y}
\ee
where $y=\sqrt{1+8 z}$. Differentiating $H(z)$ once with respect to $z$ gives
\bea
\frac{H'(z)}{H(z)} \approx \frac{8 (y+3)+4 (2 y+3) (y^2-1)+2 y (y+3)
(y^2-1) \ell }{y^2 (y^2-1) (y+3)} 
\label{logderiv}
\eea
Using the above expression in (\ref{zs1}) for large $y$, we get $y_s
\approx 4 J/\ell$ and therefore 
\be
z_s \approx \frac{2 J^2}{\ell^2}
\ee
On differentiating (\ref{logderiv}) once, we have
\be
\left(\frac{H'(z)}{H(z)}
\right)' \approx \frac{4}{y} \left[ \frac{4}{3
    (y+3)^2}+\frac{4}{(1+y)^2}-\frac{4+ 6 \ell}{3
    y^2}+\frac{8}{y^3}-\frac{4}{(1-y)^2}  \right]
\label{log2}
\ee
Using (\ref{logderiv}) and (\ref{log2}) in (\ref{fluct1}), we
obtain
\bea
K''(z_s) &\approx& \frac{8 \left[-36+ 6 y_s (y_s^2-3)+ y_s (y_s+3)^2
    (1+y_s)^2 \ell \right]}{y_s^4 (y_s+3)^2 (y_s^2-1)} \\
&\approx& \frac{8 \ell}{y_s^3}= \frac{a^4}{8 J^{3}}
\eea
Thus we have
\be
Q_J \approx \frac{2 J^{3/2}}{\sqrt{\pi} \ell^2} \times \frac{2
  -\alpha_-(z_s)}{\alpha_+ (z_s)- \alpha_-(z_s)} \times \frac{(1-{\tilde
    f})^{1+\alpha_-(z_s)}}{z_s^J}
\label{QJuni_expr}
\ee
where $\alpha_{\pm}$ is given by (\ref{alphapm}). 

%===============================================================

%===============================================================

\begin{figure}[ht]
\begin{center}
\includegraphics[width=0.6 \linewidth,angle=0]{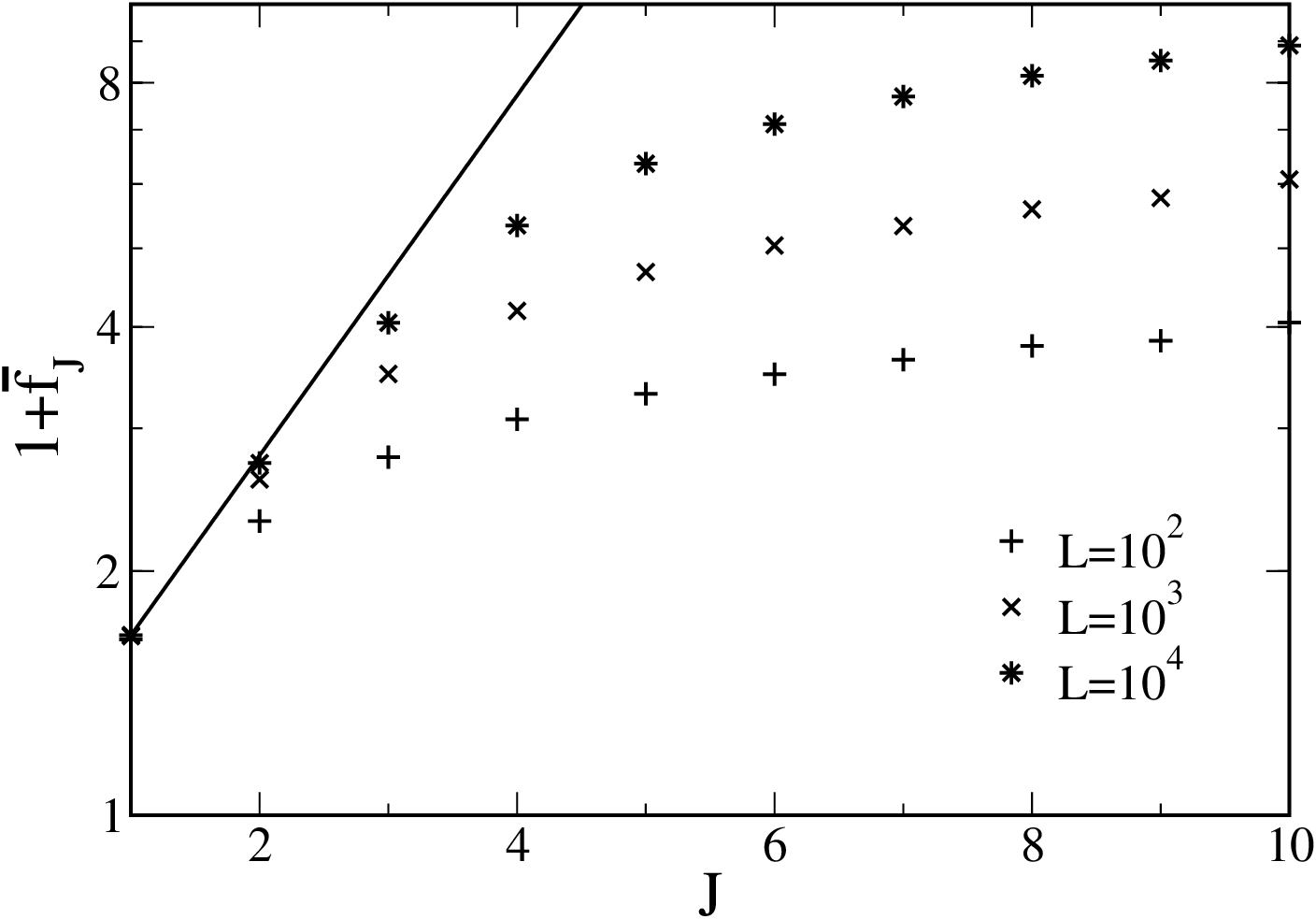}
\end{center}
\end{figure}

\begin{figure}[ht]
\begin{center}
\includegraphics[width=0.6 \linewidth,angle=0]{exp_FJ.eps}
\end{center}
\end{figure}

\end{document}